\documentclass[oldversion]{aa}

\usepackage{graphicx}
\usepackage{booktabs}
\usepackage{array}
\usepackage[stable]{footmisc}
\usepackage{subcaption}
\usepackage{graphicx}
\usepackage{natbib}
\newcolumntype{C}[1]{>{\centering\let\newline\\\arraybackslash\hspace{0pt}}m{#1}}

\usepackage{txfonts}

\usepackage{hyperref}

\begin{document} 

\title{Web application for galaxy-targeted follow-up of electromagnetic counterparts to gravitational wave sources}

\author{L.~Salmon\inst{1} \and L.~Hanlon\inst{1} \and R.~M.~Jeffrey\inst{1} \and A.~Martin-Carrillo\inst{1}}
  
\institute{School of Physics, University College Dublin,
            Belfield, Dublin 4, Ireland \\ \email{lana.salmon@ucdconnect.ie}}

\date{Received 26 August 2019 / Accepted 11 December 2019}

  \abstract
  {The Laser Interferometer Gravitational Wave Observatory (LIGO) and Virgo Collaboration's Observing Run 3 has demanded the development of widely-applicable tools for gravitational wave follow-up. These tools must address the main challenges of the multi-messenger era, namely covering large localisation regions and quickly identifying decaying transients.
   To address these challenges, we present a public web interface to assist astronomers in conducting galaxy-targeted follow-up of gravitational wave events by offering a fast and public list of targets post-gravitational wave trigger.
   After a gravitational wave trigger, the back-end galaxy retrieval algorithm identifies and scores galaxies based on the LIGO and Virgo computed probabilities and properties of the galaxies taken from the Galaxy List for the Advanced Detector Era (GLADE) V2 galaxy catalogue. Within minutes, the user can retrieve, download, and limit ranked galaxy lists from the web application. The algorithm and website have been tested on past gravitational wave events, and execution times have been analysed. The algorithm is being triggered automatically during Observing Run 3 and its features will be extended if needed. The web application was developed using the Python based \texttt{Flask} web framework. The web application is freely available and publicly accessible at gwtool.watchertelescope.ie.
}


  \maketitle

\section{Introduction}

The detection of GW170817 by the Advanced Laser Interferometer Gravitational Wave Observatory (LIGO) \citep{aasi2015advanced} and the Advanced Virgo detector \citep{acernese2014advanced} triggered an international campaign of electromagnetic (EM) follow-up by ground and space-based observatories. This follow-up led to the detection of the almost coincident short-lived gamma-ray emission \citep{goldstein2017ordinary,savchenko2017integral} and subsequent longer kilonova emission $\sim$2\,kpc from the galaxy NGC4993 \citep{abbott2017multi,andreoni2017follow,arcavi2017optical,coulter2017swope,cowperthwaite2017electromagnetic,drout2017light,evans2017swift,kilpatrick2017electromagnetic,mccully2017rapid,pian2017spectroscopic,smartt2017kilonova,tanvir2017emergence,valenti2017discovery}. 

The gravitational wave (GW) event is consistent with the signature expected from the inspiral and merger of two neutron stars (NS), accompanied by a gamma-ray burst (GRB). Neutron star mergers are thought to be the progenitors of short GRBs ($<$2s) as some short GRBs have been found in galaxies lacking in star formation \citep{berger2014short}. The subsequent, more isotropic, transient (kilonova) is associated with radioactive decay of neutron rich material in the sub-relativistic outflow \citep{metzger2017kilonovae}.  Kilonova signatures have been found in the light curves of a handful of short GRBs (for a review, see \citet{ascenzi2019luminosity}), further strengthening their link to NS-NS mergers. This exciting event marked the end of Advanced LIGO's Observing Run 2 (O2) \citep{abbott2019low}, which was followed by technical upgrades to both LIGO and Virgo in order to improve the sensitivity and range for Observing Run 3 (O3).

Even with the enhanced capabilities of the LIGO and Virgo Collaboration (LVC) during O3, which began in April 2019, the expected rate of NS-NS mergers that are detectable by the observatory (i.e. within about 170\,Mpc) is only between one and 50 events per year \citep{abbott2018prospects}. The rate of neutron star-black hole (NS-BH) merger events, which are also expected to produce EM counterparts, is highly uncertain but likely to be even lower \citep{abbott2018prospects}. 

With so few candidates expected, we need tools to identify electromagnetic counterparts as quickly as possible. This will allow us to probe the early-time physics of these events and answer open questions about the properties and progenitors of short GRBs  \citep{d2018evolution,lamb2018grb,ziaeepour2018prompt,zhang2018peculiar}, the neutron star equation of state \citep{abbott2018gw170817nseq,annala2018gravitational,bauswein2017neutron,10.1093/mnrasl/slz133,kiuchi2019revisiting,radice2018gw170817,raithel2018tidal}, r-process nucleosynthesis \citep{drout2017light,kasen2017origin,pian2017spectroscopic}, and cosmology \citep{abbott2017ligohubble,hotokezaka2018hubble,vitale2018measuring}.

It is not simple to identify an EM counterpart to a merger event: the localisation region inferred from the GW signal is large (typically $\sim$20--1000\,deg$^2$\citep{abbott2018prospects}). Since the signal from the counterpart decays rapidly, the counterpart is often too faint to detect unless it can be localised quickly. Wide field of view telescopes (FoV$>$\,1 deg$^2$) use tiling strategies to optimise coverage of the localisation regions \citep{andreoni2019growth,coughlin2019growth,dobie2019optimised,goldstein2019growth}; however, these regions are often too large for small field of view telescopes to survey quickly.

This paper presents a public web interface to a galaxy retrieval algorithm, which combines localisation information from LIGO/Virgo with galaxy positions from the Galaxy List for the Advanced Detector Era (GLADE; \citet{dalya2018glade}) source catalogue to provide the community with a ranked list of candidate galaxies for rapid follow-up observations after a GW detection. The paper is arranged as follows. Section \ref{Sec:2} reviews current gravitational wave counterpart search strategies for optical and near-IR telescopes. Section \ref{Sec:algorithm} outlines the implementation and outputs of a galaxy ranking algorithm. The front-end web application which hosts the outputs of this algorithm is detailed in Sect. \ref{Sec:4}. Section \ref{Sec:5} presents a discussion of the results of testing, use-cases, and future development.

\begin{table*}

\caption[]{Features of existing and future wide and narrow field telescopes. The number of pointings is calculated based on the coverage of a square 10$\times$10\,deg$^2$ region with tiles that don't overlap. * denotes future telescopes. }
\label{table:comparetele}
\centering

    \begin{tabular}{p{0.2 \linewidth}p{0.15 \linewidth}p{0.15 \linewidth}p{0.15 \linewidth}p{0.15 \linewidth}p{0.03 \linewidth}}
    \hline
    \noalign{\smallskip}
    \textbf{Telescope} & \textbf{\# Apertures} & \textbf{Aperture \newline (m)}	 & \textbf{Field of View (per telescope) \newline (deg$^2$)} & \textbf{\# Pointings for 100\,deg$^2$} & \textbf{Ref} \\
    \noalign{\smallskip}
    \hline
    \noalign{\smallskip}
    \multicolumn{4}{l}{\textbf{Wide field telescopes}}\\
    \hline
    \noalign{\smallskip}
    ASSASN-1    &   4   &  0.14     & 4.5       &23     & 1 \\
    ATLAS       &   2   &  0.5      & 29        & 4     &2 \\
    DECam       &   1   &  4        & 3         & 34    &3 \\
    Evryscope	&   27  &  0.061    & 8660      & N/A   &4\\
    GOTO        &   4   &   0.4		& 5         & 20    & 5 \\
    MASTER	WFC &   16	& 0.082     & 24        &5      & 6  \\
    PANSTARRS-1	&   1   &   1.8	    & 7         & 15    &7 \\
    BlackGEM*	&	4   &   0.6     & 2.7       & 38    &8\\
    LSST*       &	1	&   8.4     & 9.6       &11     &9\\
    ZTF         &	1	&   1.2	    & 47        &3      &10 \\ \hline
    \noalign{\smallskip}
    \multicolumn{4}{l}{\textbf{Narrow field telescopes} } \\
    \hline
    \noalign{\smallskip}
    BOOTES              & 5     & 0.6   &0.16   & 625   &11 \\
    Liverpool Telescope & 1	    & 2     &0.076  &  1315 &12 \\
    Watcher             & 1     & 0.4   & 0.16  & 625   &13 \\
    \noalign{\smallskip}
    \hline
    
    \end{tabular}

	\tablefoot{(1) \citealt{kochanek2017all}; (2) \citealt{tonry2018atlas}; (3) \citealt{depoy2008dark}; (4)  \citealt{law2015evryscope}; (5)
	\citealt{dyer2018telescope}; (6) \citealt{lipunov2017master};  (7) \citealt{denneau2013pan}; (8)   \citealt{bloemen2015blackgem}; (9) \citealt{ivezic2019lsst}; (10) \citealt{bellm2018zwicky,graham2019zwicky,masci2018zwicky}; (11) \citealt{castro2012building}; (12)  \citealt{steele2004liverpool}; (13) \citealt{french2004watcher}. 
	}     
\end{table*}

\section{Review of search strategies}\label{Sec:2}

\subsection{Strategies for wide field telescopes}
Many wide field survey telescopes and telescope networks have employed a tiling strategy, whereby their large fields of view and network capabilities allow for a large localisation region to be tiled in short times. The properties of a sample of these facilities compared to some narrow field telescopes can be seen in Table \ref{table:comparetele}. For comparison, the resolution of typical LIGO/Virgo sky maps means that each pixel has an area of between 0.003--0.013\,deg$^2$, much smaller than the field of view of both wide and narrow field facilities. Global networks such as the Global Rapid Advanced Network Devoted to the Multi-messenger Addicts (GRANDMA; \citet{antier2019first}), the Global Relay of Observatories Watching Transients Happen (GROWTH;  \citet{andreoni2019growth,coughlin2019growth}), and the Mobile Astronomical System of Telescope-Robots (MASTER; e.g. \citet{lipunov2019ligo}) must coordinate the pointing strategies of their telescopes to maximise coverage and depth, and to make use of the extensive visibility of the telescopes in their network. Optical surveys such as the Zwicky Transient Facility (ZTF;  \citet{bellm2018zwicky,graham2019zwicky,masci2018zwicky}) use a fixed grid on the sky to tile a region. This can assist candidate identification via image subtraction, but it does not optimise the coverage of a GW localisation region. The challenge faced by these observatories is to find the smallest number of tiles that can cover the entire localisation region, and/or swiftly identify the counterpart. 

The tiling and scheduling strategies developed to face this challenge can be classified broadly into three approaches. The first is the probability-ranked observation of fixed-grid tiles (e.g. \citet{coughlin2016maximizing,ghosh2016tiling}). The second approach is the iterative placing of tiles to maximise the probability covered (e.g. \citet{ghosh2016tiling}), and third is the detectability-based observation of tiles (e.g. \citet{salafia2017and}). These algorithms are outlined, compared, and implemented in the \texttt{gwemopt} code \citep{coughlin2018optimizing,coughlin2019teamwork} along with various scheduling strategies. For a review of these strategies see \citet{coughlin2018optimizing, rana2017enhanced}. 

\begin{table}
\caption{Comparison of the tiles required to cover a sample of LVC localisation regions with ZTF, ATLAS, and Watcher and the fraction of these tiles which contain galaxies. The 99\% regions are covered in non-overlapping tiles calculated using the \texttt{sky\_tiling} code. Tiles are equal in size to the telescope field of view, as stated in Table \ref{table:comparetele}. The `Galaxy Tiles' column represents the number of tiles which contain galaxies identified by the algorithm described in this paper.}

\label{table:compare}
\centering

    \begin{tabular}{p{0.17\columnwidth}|>{\centering\arraybackslash}p{0.06\columnwidth}>{\centering\arraybackslash}p{0.1\columnwidth}|>{\centering\arraybackslash}p{0.06\columnwidth}>{\centering\arraybackslash}p{0.1\columnwidth}|>{\centering\arraybackslash}p{0.1\columnwidth}>{\centering\arraybackslash}p{0.1\columnwidth}}
    \hline
    \textbf{} & \multicolumn{2}{c|}{\textbf{ZTF}}      & \multicolumn{2}{c|}{\textbf{ATLAS}}    & \multicolumn{2}{c}{\textbf{Watcher}}  \\ 
    \textbf{GW event}         & \textbf{Tiles} & \textbf{Galaxy Tiles} & \textbf{Tiles} & \textbf{Galaxy Tiles} & \textbf{Tiles} & \textbf{Galaxy Tiles} \\ 
    \hline

    S190814bv 	&	 	58		&   58  &   98  &    98 &   12\,803 &   9\,248  \\
    S190828j	&	 	44		&   42  &   72  &     69    &  7\,717 &    3\,677 \\
    S190910d	&	 	191		&   190 &   334 &   333 &   210\,290 &  40\,273  \\
    S190923y	&	 	120		&   118 &   205 &   203 &  27\,708&    23\,834\\
    S190924h	&	 	30		&   30  &   51  &   51  &   6\,567 &    6\,222 \\
    S190930s	&	 	100		&   100 &   175 &   174 &  24\,352 &   19\,512 \\
    S190930t	&	 	779		&   764 &   1\,412  &    1\,391  &  210\,419 &    61\,563 \\

            	\hline
         \end{tabular}
     	
	%tablefoot{The first column \texttt{id} is used to select the source mapping when we instantiate the class \SM\  (see in text for more details).}     
\end{table}

\subsection{Strategies for narrow field telescopes}
Tiling strategies are impractical for narrow field of view telescopes to implement due to the large LVC localisation regions. The primary motivation behind this work is to provide the astrophysics community with a public web application to support galaxy-targeted searches of LVC localisation regions, more suited to telescopes with narrow fields of view. The main advantage of a galaxy-targeted strategy is the reduction in the number of pointings required by a factor of 10--100 \citep{gehrels2016galaxy}. As an example, the 99\% localisation regions of a sample of past GW events were covered with tiles of varying size, representing the fields of view of two current survey telescopes, ZTF and the Asteroid Terrestrial-impact Last Alert System (ATLAS), and a typical robotic telescope, such as the Watcher robotic telescope.  Table \ref{table:compare} outlines the tiles required by each telescope to cover the entire region, as identified using the \texttt{sky\_tiling}\footnote[1]{\href{https://github.com/shaonghosh/sky_tiling.git}{https://github.com/shaonghosh/sky\_tiling.git}} code. For comparison, Table \ref{table:compare} also shows the subset of tiles which contain galaxies identified using the algorithm outlined in Sect. \ref{Sec:algorithm} of this paper. 

\begin{figure*}
	\centering
  		\includegraphics[width=\linewidth]{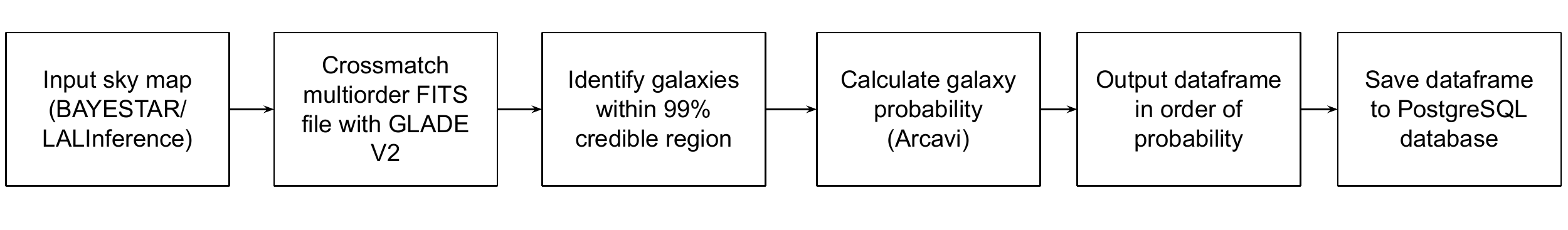}
  		\caption{Flowchart of the galaxy retrieval algorithm. The dataframes are saved as PostgreSQL database tables for each contour. This database is hosted on Amazon S3.}
  		\label{fig:flowchart}
\end{figure*}

Rapid observation of fields of interest, preferably within one night, is required due to the fast evolving counterpart. For example, the kilonova associated with GW170817 decayed at a rate of 1.1 mag/day in the r band \citep{valenti2017discovery}. The 22 galaxies identified to be within the 90\% localisation region for this source could easily have been covered in one night of observing. The smaller number of fields also presents observers with the opportunity to allocate more time to observe each field and leads to quicker identification of counterparts.

An efficient way of determining the order in which to observe galaxies is to retrieve and rank the galaxies based on known properties of those galaxies and LVC provided probabilities. Various ranking algorithms have been outlined and implemented by telescopes and networks (e.g. \citet{dobie2019optimised, ducoin2019optimising, kasliwal2017illuminating, klingler2019swift, rana2019galaxy, yang2019optical}). Section \ref{Sec:algorithm} outlines a galaxy retrieval algorithm which utilises \citet{arcavi2017optical}.

\section{Galaxy retrieval algorithm} \label{Sec:algorithm}

 The galaxy retrieval algorithm (Fig. \ref{fig:flowchart}) is implemented in Python 3.6.6 to interface with the \texttt{Flask} web framework and the \texttt{ligo.skymap} package. Figure \ref{fig:flowchart} illustrates a flowchart of the algorithm which takes as input the sky map and GLADE galaxy catalogue and outputs a ranked galaxy list ordered by the probability of an association between the GW event and the galaxy.

\subsection{Inputs}
Each GW trigger contains the source classification, source localisation, and signal quality. The probability that the event originated from each source (binary black hole (BBH),  binary neutron star (BNS), NS-BH, MassGap, Terrestrial) is used to determine the most likely merger event that occurred. Also included in the trigger is the probability that the merger contains a NS, p(HasNS), and the probability that there is non-zero remnant matter, p(HasRemnant). These parameters are used to determine the likelihood of an EM counterpart \citep{chatterjee2019machine}. The signal quality is quantified by the false alarm rate (FAR), and the localisation and distance estimates of the source are included in a sky map file \citep{singer2016supplement}. The decision to follow-up an event is driven by the parameters included in the trigger, however the galaxy retrieval algorithm outlined in this section can respond to all events regardless of their classification.
\subsubsection{LVC sky map}
To choose which part of the sky is observed, and when, the `sky map' provided by the LVC is needed. Following a GW event, the BAYESTAR algorithm \citep{singer2016rapid} or the LALInference algorithm \citep{lalsuite,veitch2015parameter} outputs a sky map made up of pixels generated using the HEALPix projection \citep{gorski2005healpix}. Each pixel in the HEALPix projection contains the following four values: the probability that the source is located in the pixel, a distance estimate, a standard deviation on the distance estimate, and a normalisation coefficient \citep{singer2016rapid}. The header of the sky map also contains posterior mean of distance (\texttt{DISTMEAN}) and standard deviation of distance (\texttt{DISTSTD}) marginalised over the whole sky. For an overview of the parameters contained in the header and pixels, see Table 1 and 2 of \citet{singer2016supplement}. The LVC sky maps can be analysed in Python using the 
\texttt{ligo.skymap} package. 

\subsubsection{GLADE galaxy catalogue}
\begin{figure}
	\centering
  		\includegraphics[width=\columnwidth]{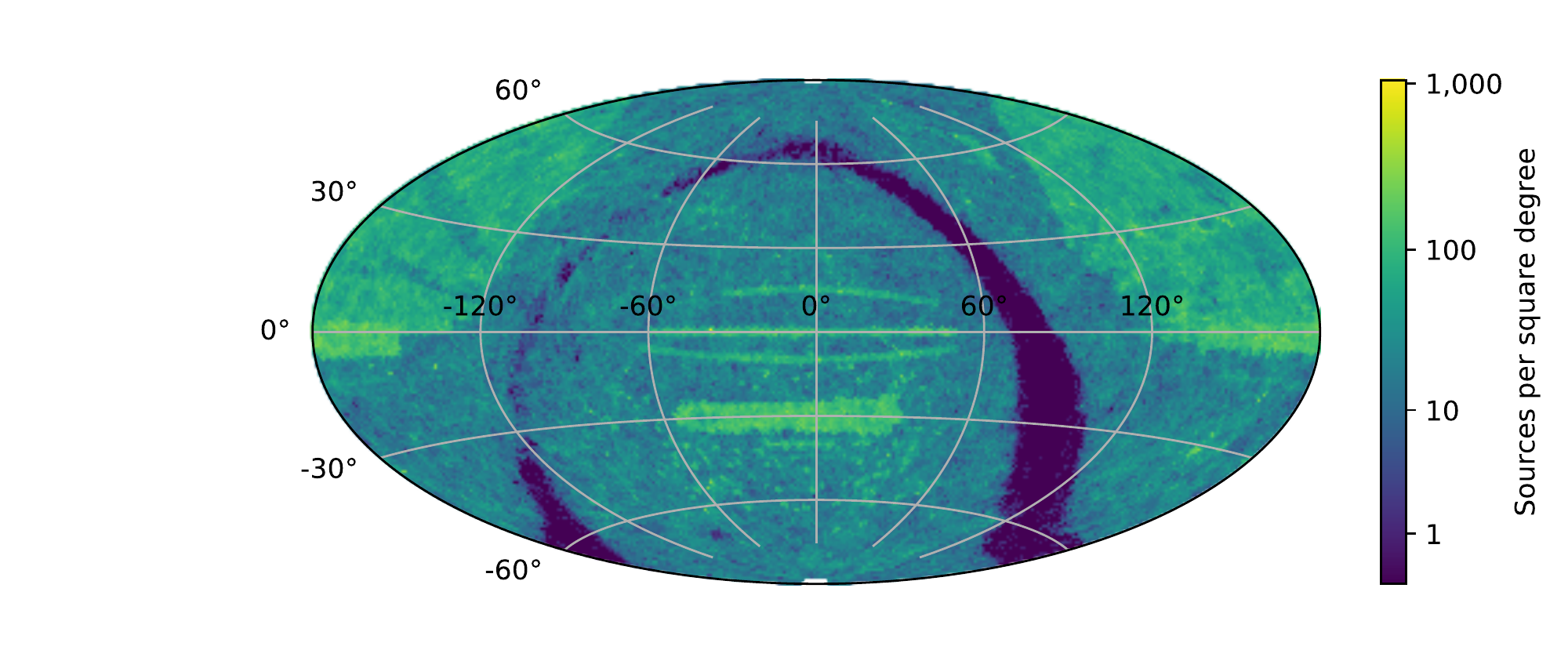}
  		\caption{Density of sources in the filtered GLADE V2 catalogue, displayed in RA and Dec. The colour scheme shows the number of galaxies per square degree; each pixel in the LVC data products has an area of between 0.003 and 0.013 square degrees. The varied coverage within the catalogue -- especially along the Galactic plane -- is obvious.   }
  		\label{fig:heatmap}
\end{figure}

In order to conduct galaxy-targeted follow-up, a galaxy catalogue is required. GLADE V2\footnote[2]{\href{http://glade.elte.hu/index.html}{http://glade.elte.hu/index.html}} \citep{dalya2018glade} contains over 3.2 million objects, with over 2.9 million of these classified as a galaxy.  It can be seen in Table \ref{tab:GLADE} that over 1.6 million of these galaxies have corresponding distance and B magnitude values. This subset is referred to as the filtered GLADE V2. A map of the filtered GLADE V2 galaxy density can be seen in Fig. \ref{fig:heatmap}. 

We choose to use the filtered GLADE V2 catalogue in this work.
When compiling GLADE V1, \citet{glade} used regression to estimate either or both of the redshift and B magnitude for over 400\,000 galaxies. GLADE V2 omits these galaxies, and adds many new galaxies. While many of the newly added galaxies do not have associated B magnitude or redshift values, we choose GLADE V2 as it includes the newest versions of its constituent catalogues, including the Two Micron All Sky Survey (2MASS; \citet{skrutskie2006two}), HyperLEDA \citep{makarov2014hyperleda}, and the Gravitational Wave Galaxy Catalogue (GWGC; \citet{white2011list}).

The completeness of the filtered GLADE V2 catalogue is close to that of the filtered GLADE V1 catalogue which omits the galaxies whose parameters were estimated via regression. Completeness distances can be seen in Table \ref{tab:GLADE}. Completeness is measured by considering the cumulative blue luminosity of the galaxies within the filtered GLADE V1 \& V2 catalogues compared to the expected blue luminosity density from a homogeneous complete galaxy catalogue given by \citet{kopparapu2008host}.
  
\begin{table*}

	\caption[]{Properties of GLADE V1 and GLADE V2. Completeness distance is measured by comparing the cumulative blue luminosity density within distinct distance limits to the luminosity density expected from a homogeneous complete galaxy catalogue, as outlined by \citet{kopparapu2008host}.}
    \label{tab:GLADE}
    \centering
    \begin{tabular}{C{0.1\linewidth}C{0.16 \linewidth}C{0.16 \linewidth}C{0.16\linewidth}C{0.16\linewidth}C{0.16\linewidth}}
    \hline
    \noalign{\smallskip}
    \textbf{Catalogue} &\textbf{\# Galaxies} & \textbf{\# galaxies without inferred distance} & \textbf{\# galaxies with known distance and B magnitude} & \multicolumn{2}{c}{\textbf{Completeness distance}}\\ 
    \textbf{} &\textbf{} &  \textbf{} & \textbf{} & \textbf{100\%} & \textbf{50\%} \\ 
    \noalign{\smallskip}
    \hline
    \noalign{\smallskip}

    GLADE V1	&	 	 1\,918\,147    &	1\,490\,234	    &   1\,490\,234	    & 37\,Mpc   & 164\,Mpc\\ \noalign{\smallskip}
    \hline
    \noalign{\smallskip}
    GLADE V2	&	 	2\,965\,718	    &   2\,965\,718	    &   1\,613\,030     & 37\,Mpc   & 148\,Mpc \\
    \noalign{\smallskip}
    \hline
    \noalign{\smallskip}
    \end{tabular}
     	
\end{table*}

\subsection{Initial galaxy sample cut}
The algorithm developed for pre-processing of sky maps has been updated for the second half of O3 (O3b) to reduce execution time. From November 2019, V1 of this algorithm is used and prior to this V0 was used. Major updates include the analysis of only the 99\% localisation region rather than the 50\%, 90\%, and 99\% regions, the increase of distance limits to $\pm$ 5\,\texttt{DISTSTD} from $\pm$ \texttt{DISTSTD} and the removal of identification of the contour each galaxy lies within. The discussion of these changes, and details of V0 of the algorithm, can be seen in Sect. \ref{section:evaluation}. V0 and V1 of the algorithm can be found on separate branches of GitHub\footnote[3]{\href{https://github.com/Lanasalmon/HOGWARTs}{https://github.com/Lanasalmon/HOGWARTs}.}.

%\subsubsection{V1 - November 2019}

\begin{figure*}[htbp!]
  \begin{subfigure}[b]{1\columnwidth}
    \includegraphics[width=1\linewidth]{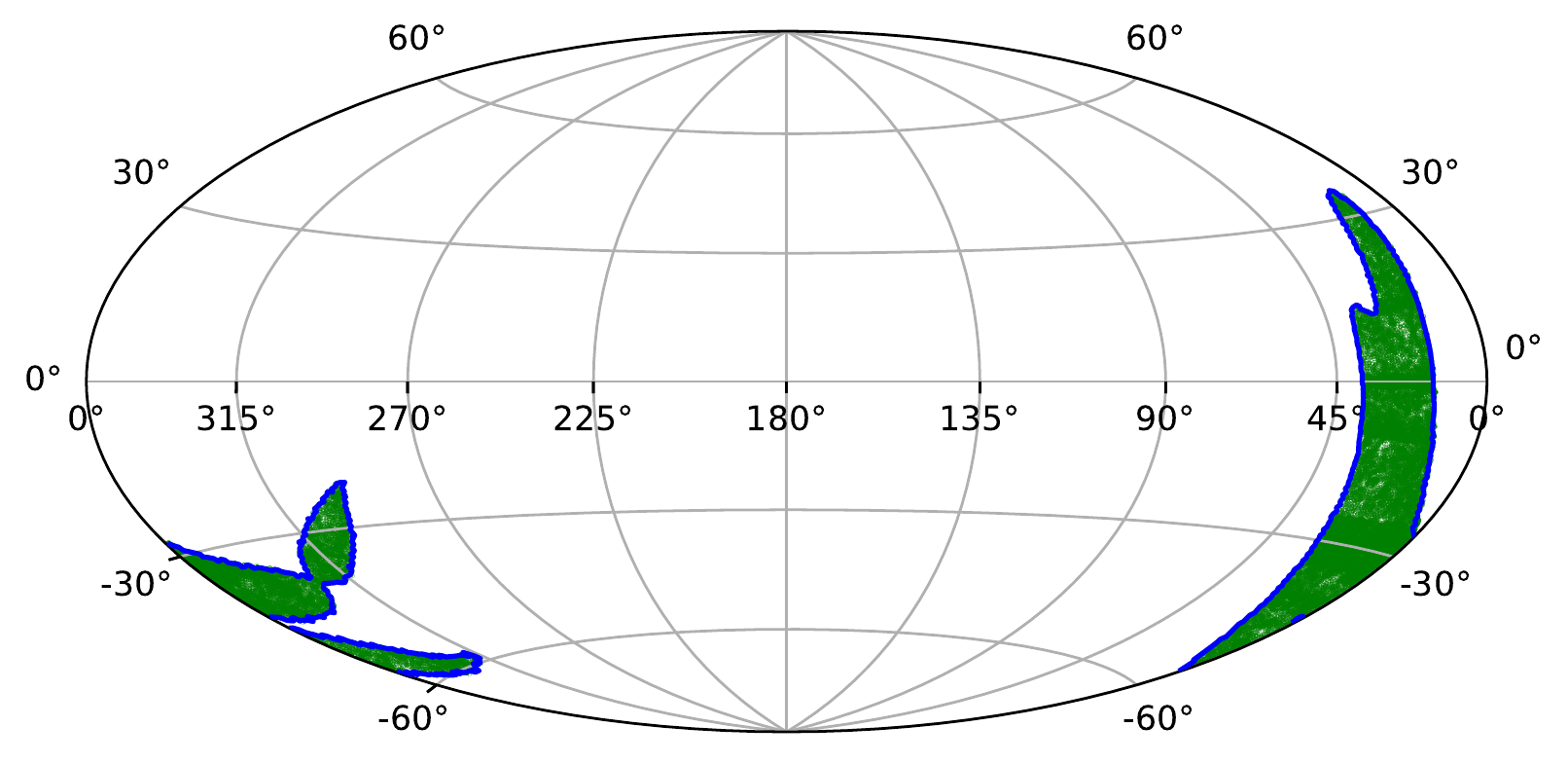}
    \caption{S190814bv}
    \label{fig:190814}
  \end{subfigure}
  \hfill %%
  \begin{subfigure}[b]{1\columnwidth}
    \includegraphics[width=\linewidth]{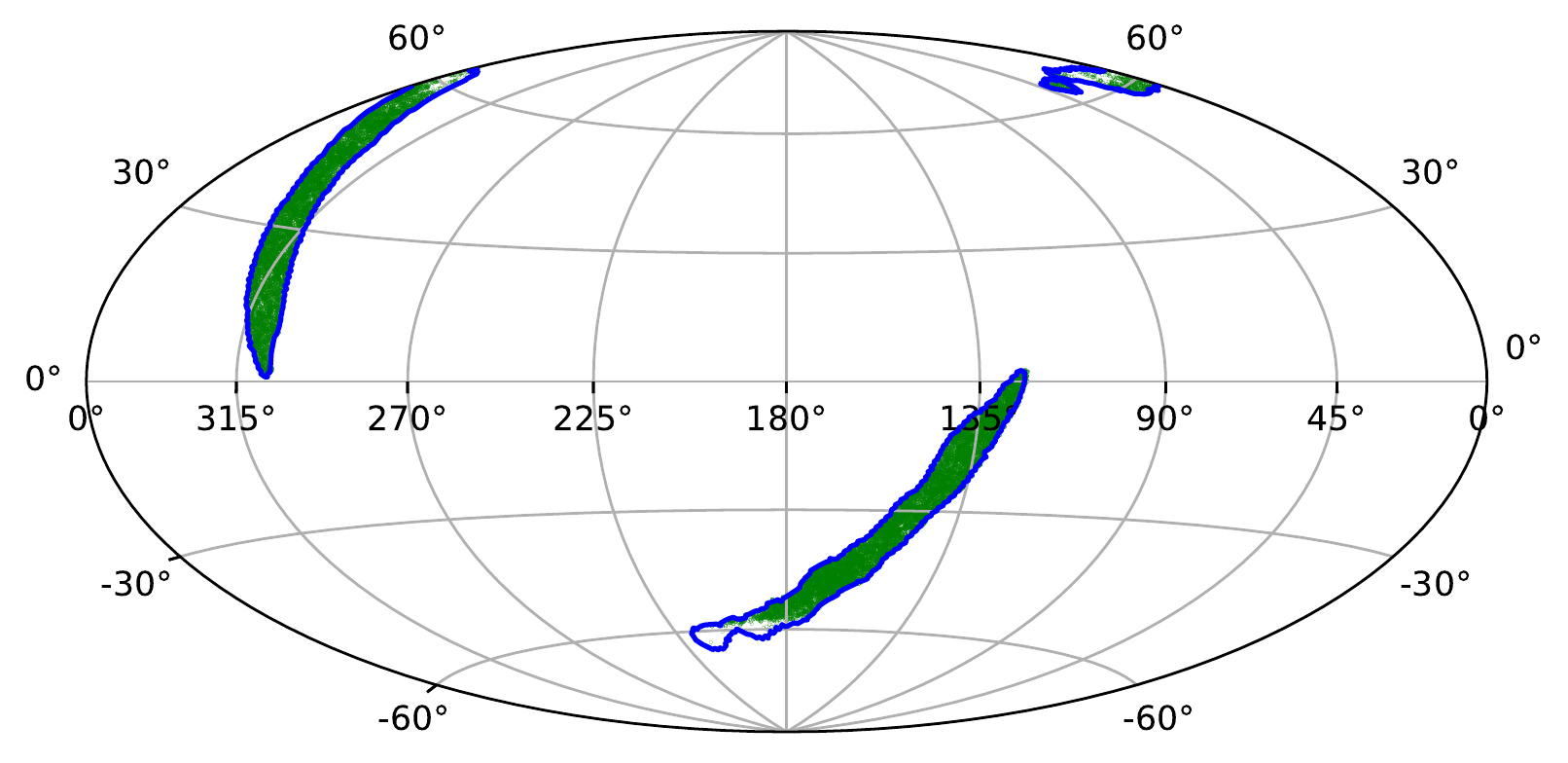}
    \caption{S190828j}
    \label{fig:190828}
  \end{subfigure}

  \begin{subfigure}[b]{1\columnwidth}
    \includegraphics[width=\linewidth]{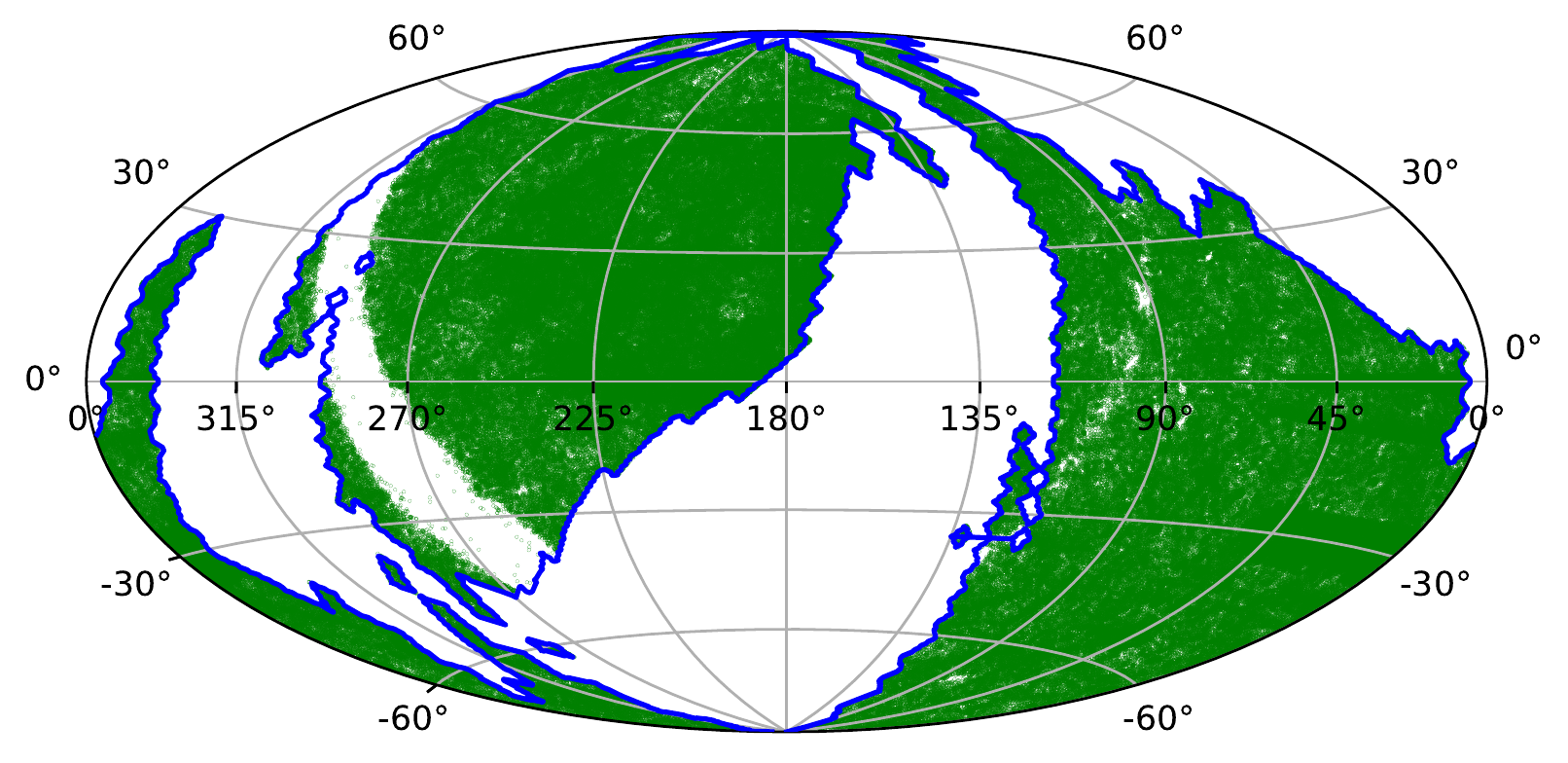}
    \caption{S190901ap}
    \label{fig:190901}
  \end{subfigure}
\hfill
  \begin{subfigure}[b]{1\columnwidth}
    \includegraphics[width=\linewidth]{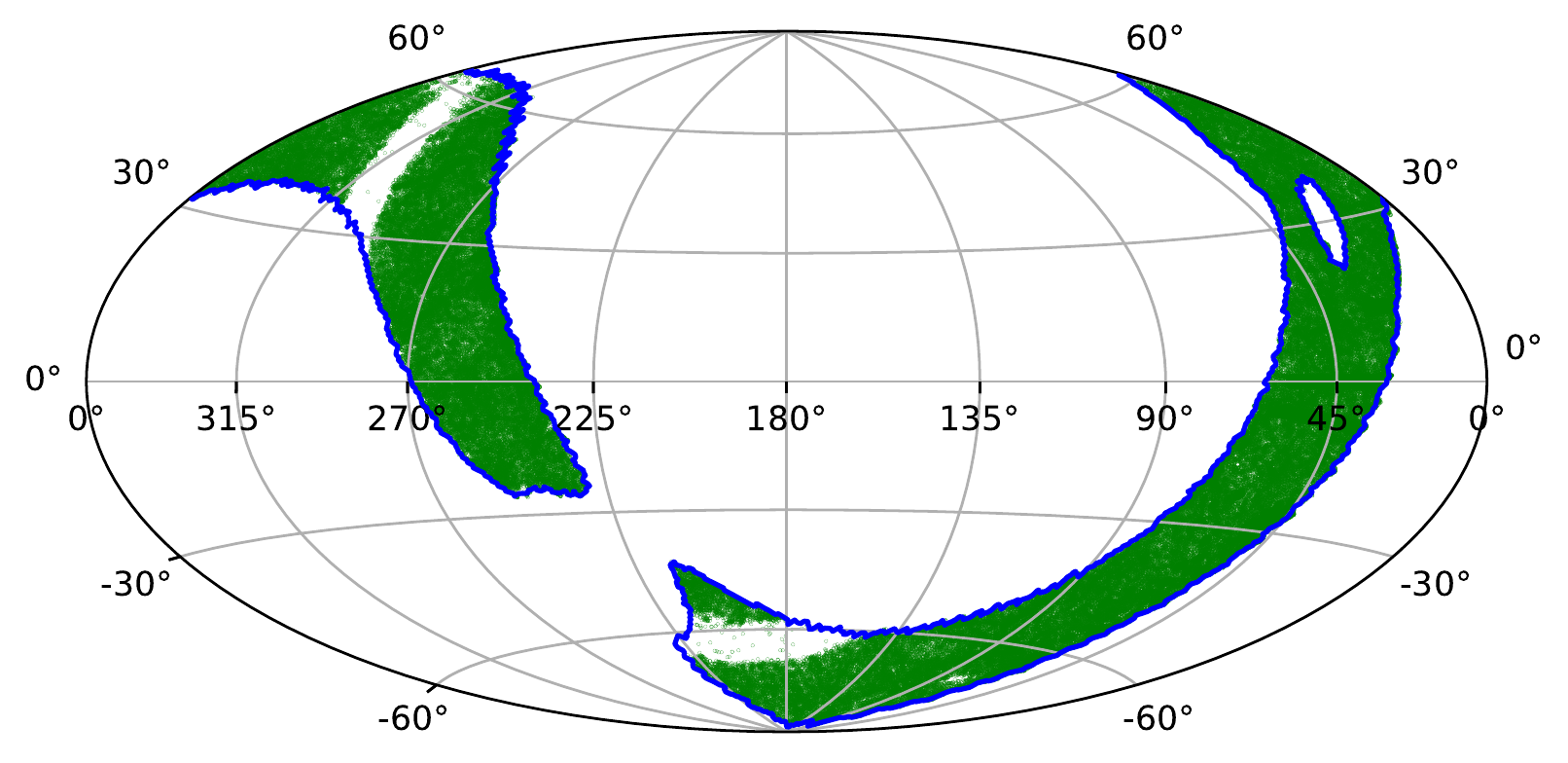}
    \caption{S190910d}
    \label{fig:190910}
  \end{subfigure}

  \begin{subfigure}[b]{1\columnwidth}
    \includegraphics[width=\linewidth]{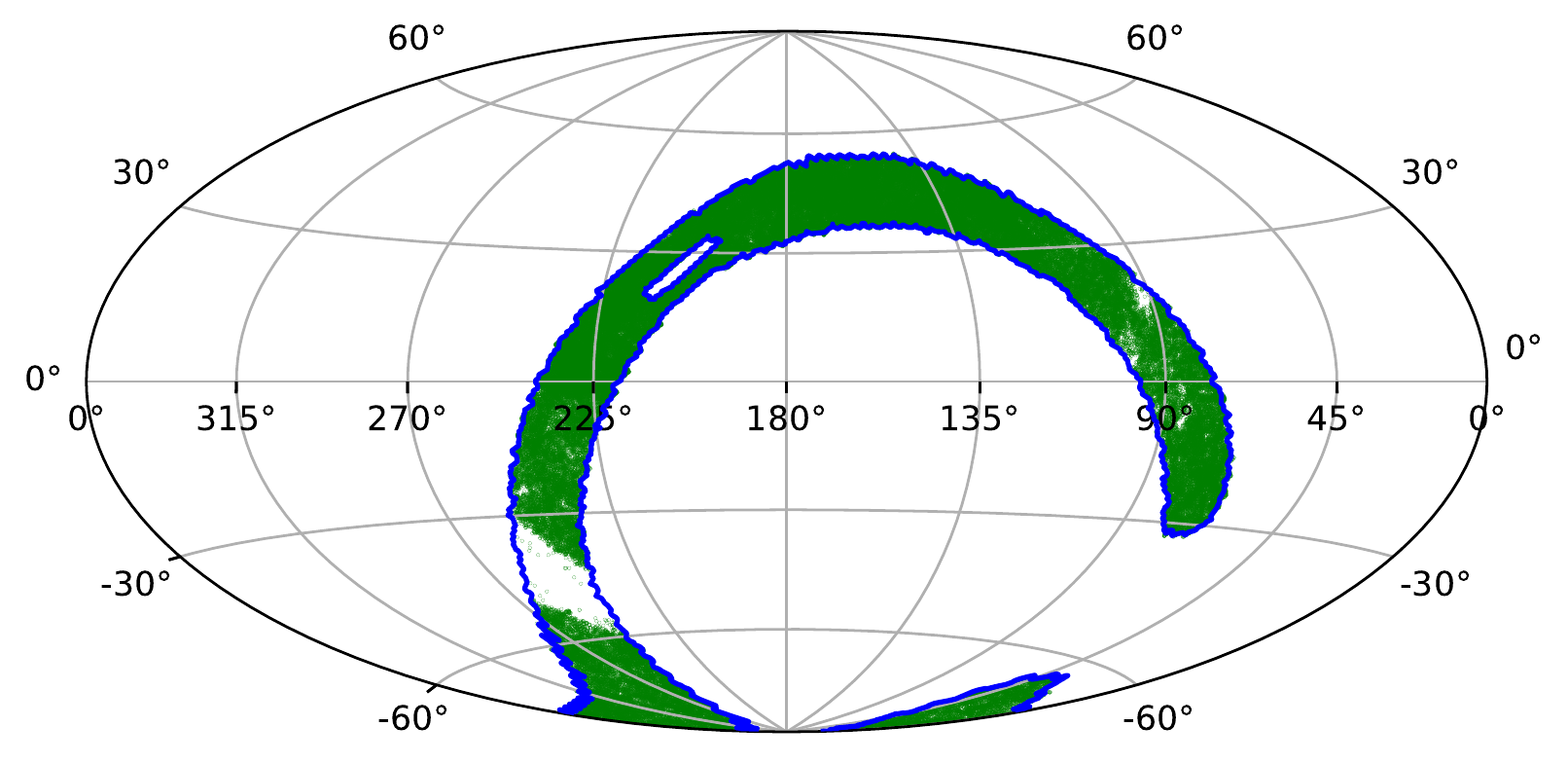}
    \caption{S190923y}
    \label{fig:190923}
  \end{subfigure}
\hfill
  \begin{subfigure}[b]{1\columnwidth}
    \includegraphics[width=\linewidth]{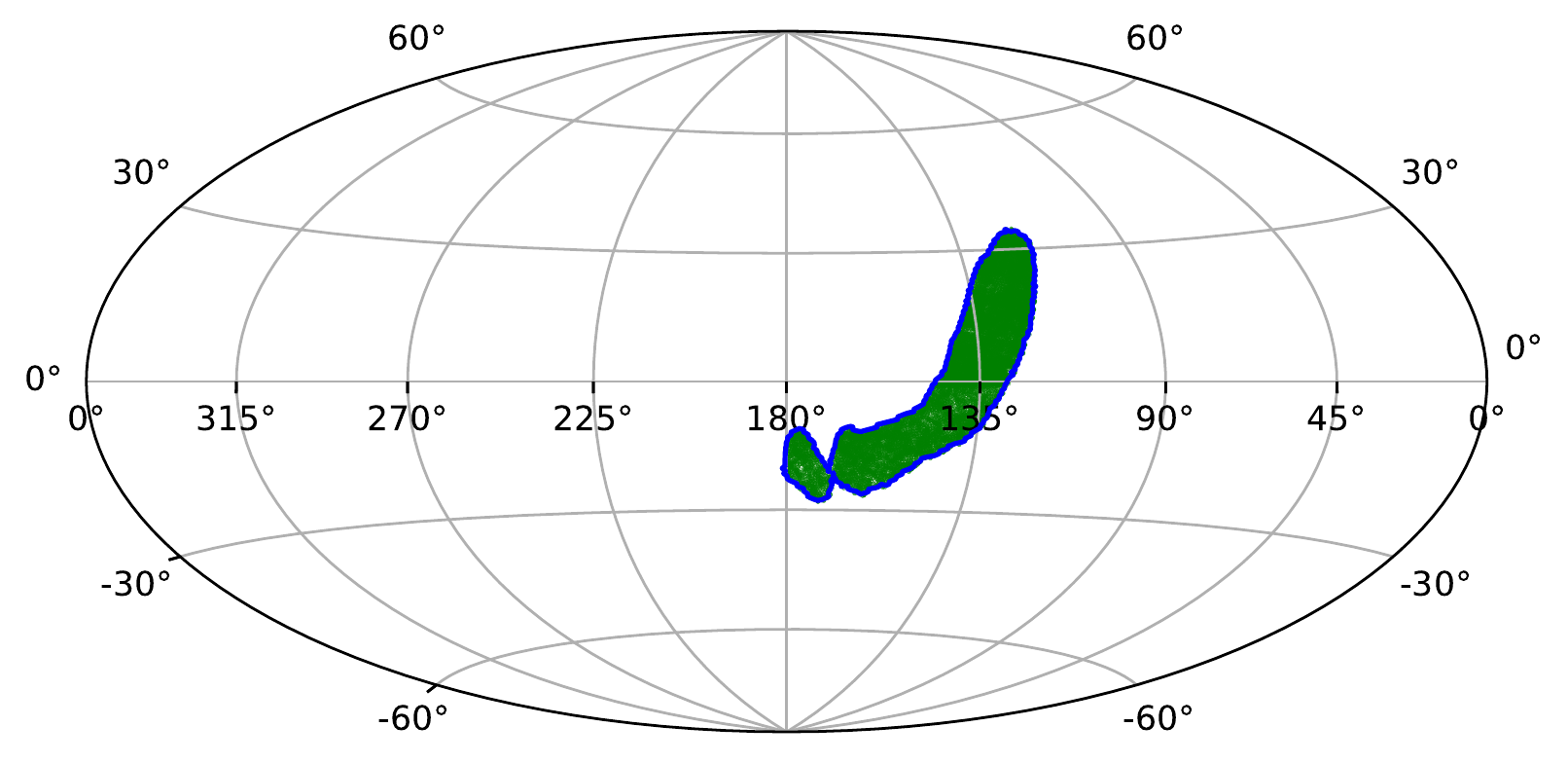}
    \caption{S190924h}
    \label{fig:190924}
  \end{subfigure}

  \begin{subfigure}[b]{1\columnwidth}
    \includegraphics[width=\linewidth]{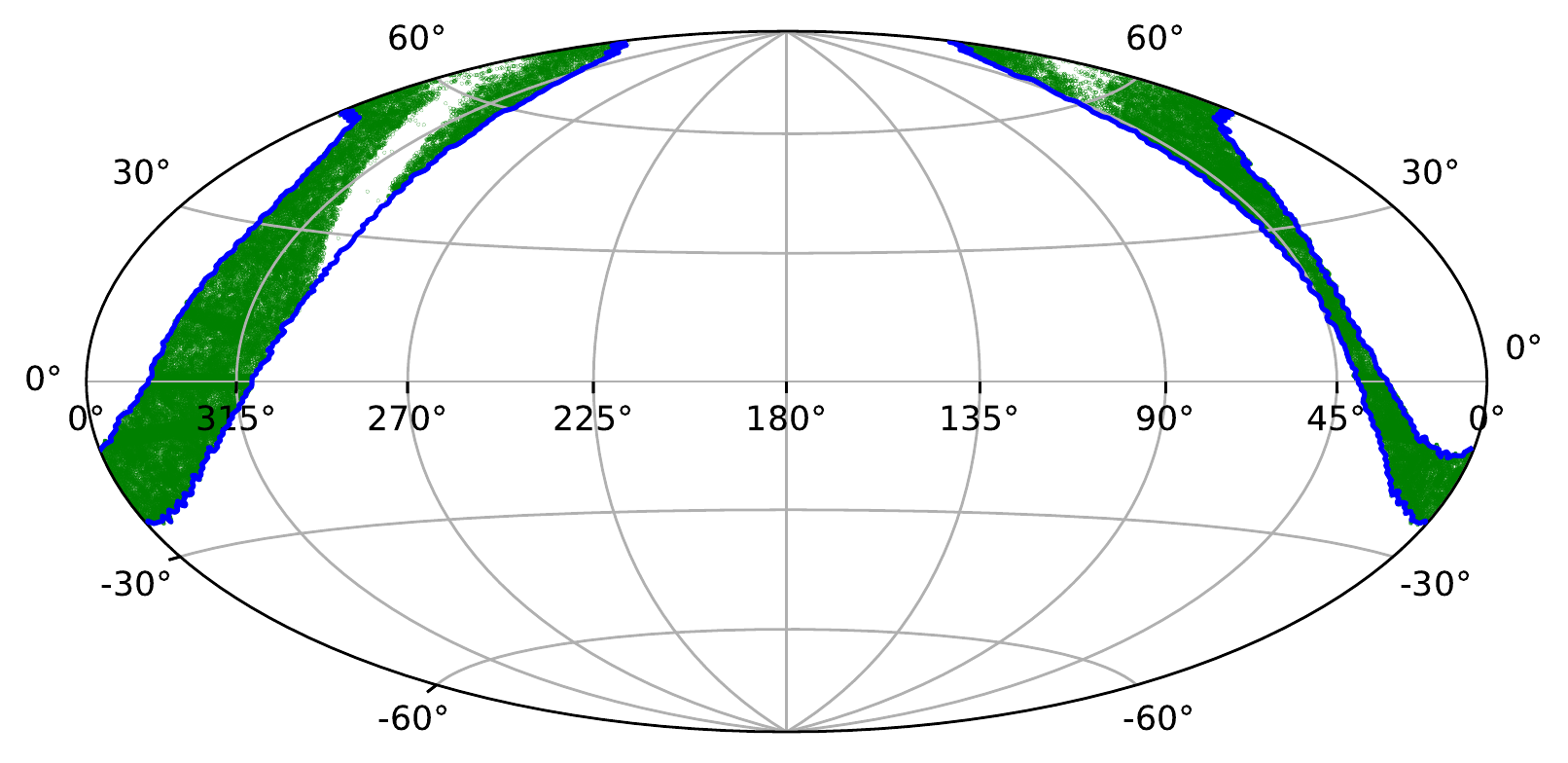}
    \caption{S190930s}
    \label{fig:190930s}
  \end{subfigure}
\hfill
  \begin{subfigure}[b]{1\columnwidth}
    \includegraphics[width=\linewidth]{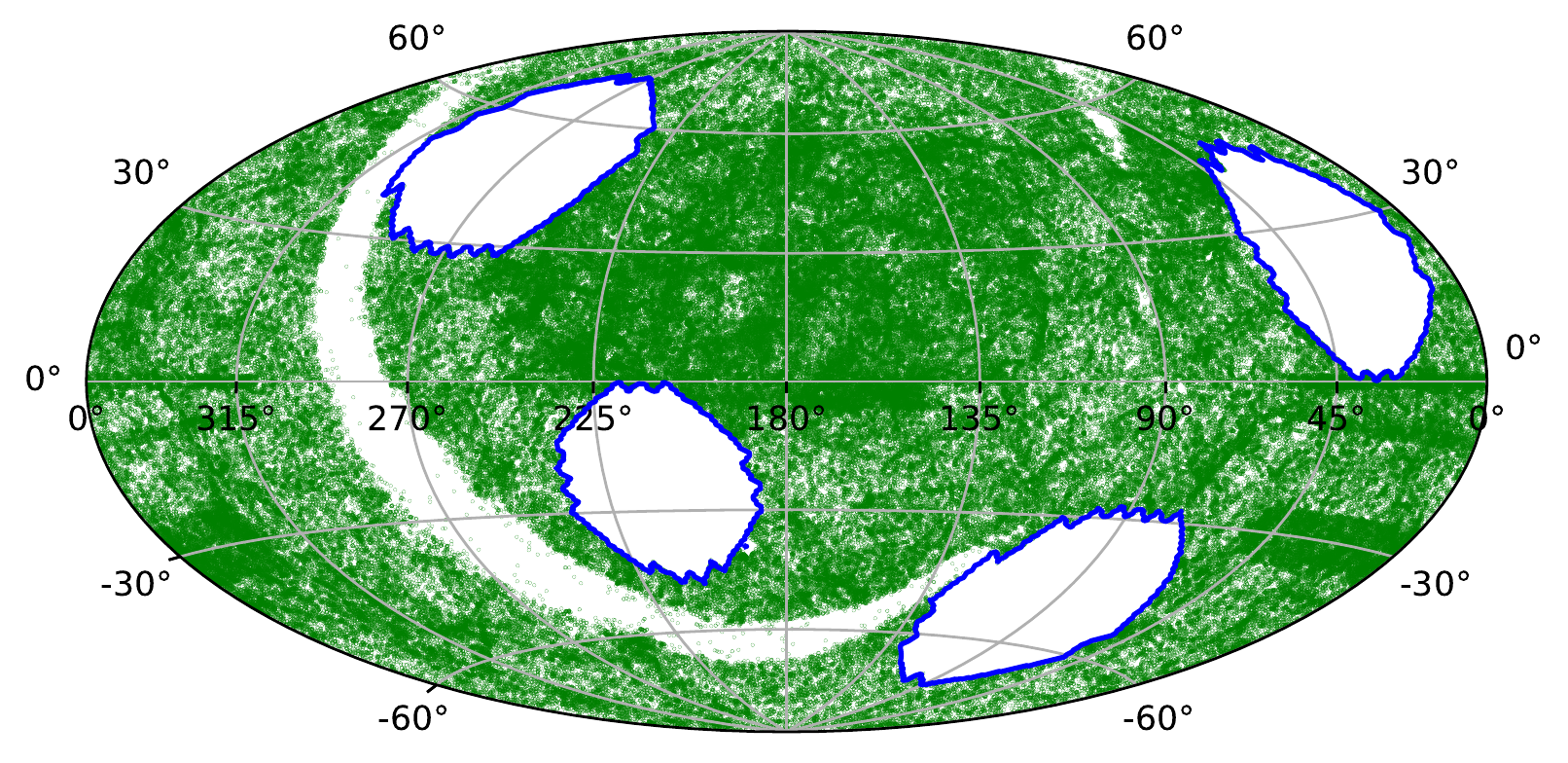}
    \caption{S190930t}
    \label{fig:190930t}
  \end{subfigure}

  \caption{99\% contour regions (blue lines) and galaxies from the filtered GLADE V2 catalogue (green points) which lie within these regions for a sample of initial sky maps for GW events from O3. We note that the galaxies are filtered based on the LVC distance estimate and error, within $\pm$ 5\,\texttt{DISTSTD}. These localisation regions are often irregularly shaped and have varied sizes. The uneven coverage of the filtered GLADE V2 catalogue is evident in these figures, especially in Fig. \ref{fig:190930t}. The galaxy density in each sky map is different due to the differing distance estimates (and completeness of the galaxy catalogue to that distance) and the magnitude of the corresponding standard deviations on distance.} 
    \label{fig:skymaps}
\end{figure*}

In V1 of the algorithm, \texttt{ligo.skymap} is used for sky map I/O and also facilitates the crossmatching of a local version of the GLADE galaxy catalogue, stored as a hdf5 file, with the sky map. The galaxies within the 99\% localisation regions and within \texttt{DISTMEAN} $\pm $ 5\,\texttt{DISTSTD} are identified using the \texttt{ligo.skymap.crossmatch} function. The coordinates, B magnitudes, and distances to these galaxies, as taken from the filtered GLADE V2 catalogue, are stored in an array to be used in subsequent probability calculations. Figure \ref{fig:skymaps} shows the galaxies from the filtered GLADE V2 catalogue identified to be within the 99\% localisation regions for a variety of past GW events.

\subsection{Galaxy ranking}
 The probability of association of the GW source with a given galaxy is calculated based on the approach outlined by \citet{gehrels2016galaxy} and the prioritisation algorithm described by \cite{arcavi2017optical}, as follows:

\begin{enumerate}
\item The location probability measure is given as
\begin{equation}
S_{loc}=p_{loc}\,p_{dist}.
\end{equation}
The probability that the GW source is at a certain location, $p_{loc}$, is obtained from the pixel at the position of the galaxy in the sky map. \\
\\
The distance to the merger computed by the LVC is contained in the pixel at the position of the galaxy. It is compared to the distance of the galaxy extracted from the filtered GLADE V2 galaxy catalogue to calculate the distance probability measure $p_{dist}$:
\begin{equation} 
p_{ dist}=N_{ dist}\,\exp \left(\frac{-[D-\mu_{ dist}]^2}{2\sigma_{ dist}^2}\right),
\end{equation}
where $N_{dist}$ is a normalising factor, $\mu_{dist}$ is the distance estimate, and $\sigma_{dist}$ is the distance error computed by the BAYESTAR/LALInference algorithms and contained in the pixel at the galaxy's position sky map. D is distance to the galaxy from the filtered GLADE V2 catalogue. 

\item Short GRBs are found in the most massive galaxies, and B luminosity is a proxy for galaxy mass \citep{berger2014short}. Brighter galaxies are assigned a larger probability. The B-band luminosity is calculated using the apparent B magnitude and distance from the filtered GLADE V2 catalogue. This is used to calculate the luminosity probability measure $S_{lum}$:
\begin{equation}
S_{lum}=\frac{L_{B}}{\sum L_{B}}.
\end{equation}
\item The overall probability of the merger occurring in a galaxy is given by 
\begin{equation}
S=S_{loc}\,S_{lum}.
\end{equation}
This probability is calculated for all galaxies and then a score is computed by normalising the probabilities to add to 1.
\end{enumerate}

\subsection{Outputs}
The list of ranked galaxies is stored in a \texttt{pandas} dataframe which contains the name, Right Ascension, Declination, distance, B magnitude, probability score, and cumulative score associated with each galaxy. The dataframes are saved as tables in a PostgreSQL database hosted on Amazon S3, where the database is accessible by the website via the Heroku command line tools.

\section{HOGWARTs web application}\label{Sec:4}
\begin{figure*}[htbp!]
	\centering
  		\includegraphics[width=0.7\linewidth]{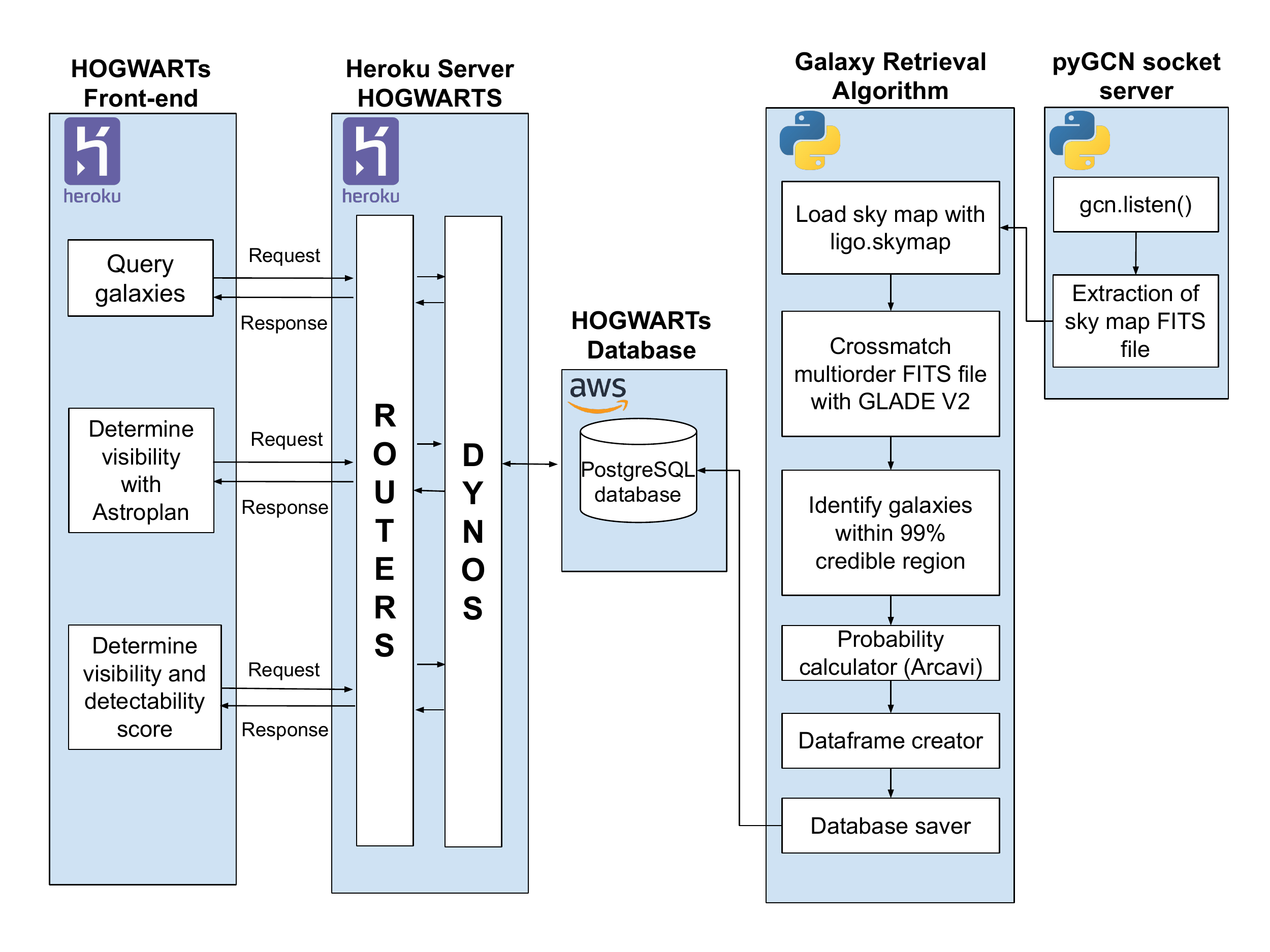}
  		\caption{Architecture of the HOGWARTs system. The galaxy retrieval algorithm saves the ranked galaxy lists for the 99\% localisation region in a PostgreSQL database on Amazon S3. This database is connected to the HOGWARTs front-end on Heroku using the Heroku API. When a user requests the galaxy list, the website queries the database via Heroku routers and dynos to obtain the list of galaxies to render. Actions may be performed on this list in the Heroku back-end, including limiting the list based on visibility or detectability. }
  		\label{fig:arch}
\end{figure*}
The HOGWARTs front-end is a free and public web application which makes the outputs of the galaxy retrieval algorithm outlined in Sect. \ref{Sec:algorithm} publicly available. The web application also allows for further operations to be performed on the outputs -- for example, the galaxy list can be limited based on the visibility in a user's location, or on user-specified limiting magnitude. 

To ensure fast and easy distribution of the outputs of the galaxy-ranking algorithm, a public web application was chosen. The only requirement for the user is a modern web browser with HTML-5 components and Javascript. This allows for universal instant access with little-to-no software installation. The distribution as a web application also allows for use by collaborations many of which have been established to tackle kilonova detection and identification during O3 using telescopes and networks. For example, to date the ElectromagNetic counterparts of GRAvitational wave sources at the VEry Large Telescope (ENGRAVE) collaboration, the GRAvitational Wave Inaf TeAm (GRAWITA; \citet{TNG2,TNG}), the Gravitational Waves at the William Hershel Telescope collaboration (GW@WHT;  \citet{WHT}), and the the GROWTH collaboration \citep{LT}  have made use of the tool to conduct follow-up observations. 

The HOGWARTs web application is implemented using the \texttt{Flask} web framework and is hosted on the cloud platform Heroku. The architecture of the HOGWARTs system can be seen in Fig. \ref{fig:arch}. \texttt{Flask} is a web framework which allows for web applications to be written in Python. \texttt{Flask} was chosen due to the flexibility of the framework. Heroku was chosen due to the ability to communicate via the command line with Amazon S3, GitHub, and the PostgreSQL database containing the results of the galaxy retrieval algorithm. 

The back-end galaxy retrieval implementation uses \texttt{pyGCN} to act as a socket listening for GW alerts from the NASA Gamma-Ray Coordinates Network (GCN)\footnote[4]{\href{https://gcn.gsfc.nasa.gov/lvc.html}{https://gcn.gsfc.nasa.gov/lvc.html}} system. When an alert occurs, the algorithm immediately analyses the sky map and creates the database tables of galaxies corresponding to the 99\% localisation regions. These are saved on Amazon S3 and the algorithm updates the website to include a new menu option for this event. 

\subsection{Inputs}

The HOGWARTs front-end web application renders the results of the galaxy retrieval algorithm in different ways depending on the user's requirements. The HOGWARTs web application currently supports three options with the possibility of adding more in the future. These are:
\begin{enumerate}
\item \textbf{Retrieve the galaxy list (Fig. \ref{fig:simplepage})}: The user chooses a GW source (e.g. GW170817) and a percentage localisation region (99\%, 90\%, or 50\%). 
\begin{figure}[htbp!]
	\centering
  		\includegraphics[width=\columnwidth]{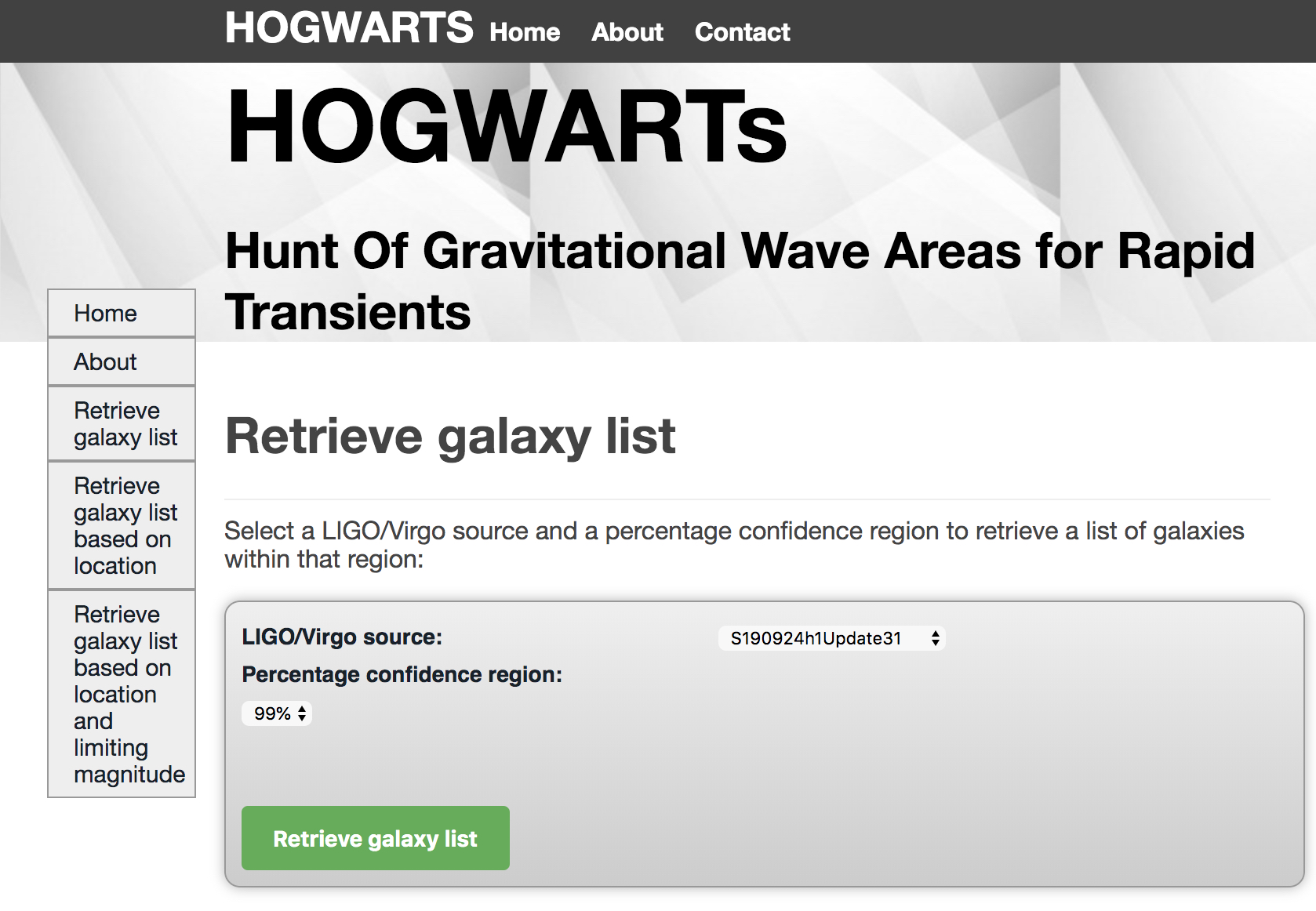}
  		\caption{`Retrieve Galaxy List' input webpage at \href{https://gwtool.watchertelescope.ie/retrieve_galaxies}{gwtool.watchertelescope.ie/retrieve\_galaxies}. The user chooses the GW source and percentage confidence region (99\%) to return the list of galaxies within those regions and a map of those galaxies. }
  		\label{fig:simplepage}
\end{figure}
\item \textbf{Retrieve only the galaxies within a region that are visible from a specific location at a user-specified time}: The user chooses a GW source (e.g. GW170817), a percentage localisation region (99\%, 90\%, or 50\%), a longitude, latitude, limiting elevation, and time of observation. Alternatively, the user can choose their observatory from the predefined list (Boyden Observatory, La Palma, Paranal, and La Silla) if it is present. \texttt{Astroplan} \citep{astropy:2018} is used to determine the galaxies visible to the user from astronomical twilight.

 \item \textbf{Retrieve only the galaxies that are visible from a specific location at a user-specified time including a detectability indicator}: The user chooses from a menu as in option 2, and also specifies the limiting magnitude.  Alternatively, the user can choose their observatory from the predefined list as in option 2. An additional column indicates the detectability of a kilonova at that distance by comparing the minimum detectable source luminosity at that distance with that of a kilonova (M$_{KNmin}$ = $-$17) at the same distance.
\end{enumerate}

\subsection{Outputs}
\begin{figure}[htbp!]
	\centering
  		\includegraphics[width=\columnwidth]{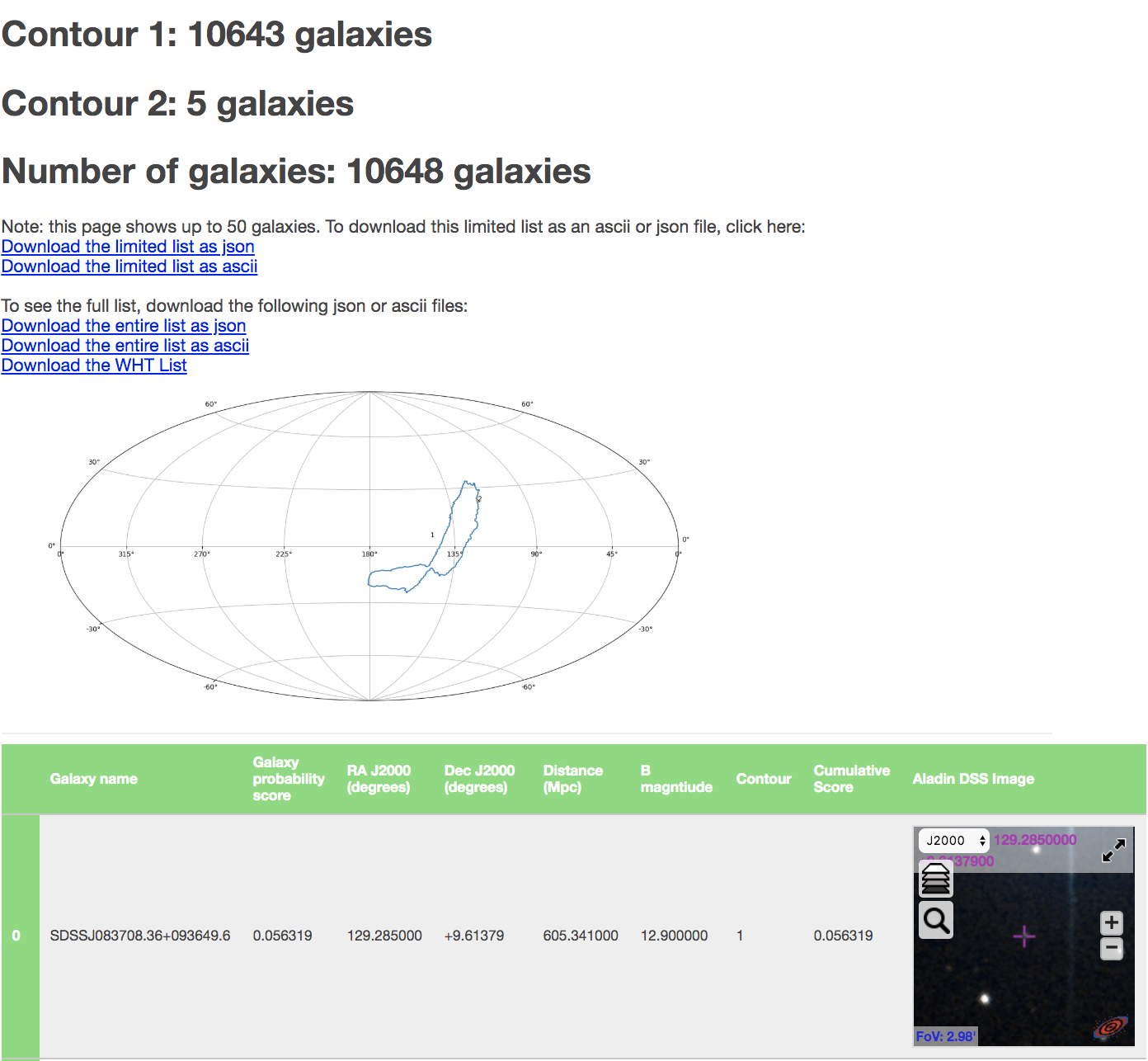}
  		\caption{Table rendered on the results webpage for the 99\% confidence region for the updated sky map of S190924h. The tables on the results page are also accompanied by a map of contours, galaxies, and the option to download the tables as json or ascii files. }
  		\label{fig:outputs}
\end{figure}

 The results webpage renders a table, ordered by probability score, containing a maximum of 100 galaxies to ensure rendering occurs in a timely manner. The table presents the galaxy name (from the GLADE V2 catalogue), probability score, Right Ascension, Declination, distance, B magnitude, contour (if V0 of the algorithm was used), cumulative probability score, and an interactive Aladin DSS image of each galaxy, as shown in Fig. \ref{fig:outputs}. A map of contour regions is also plotted. The HOGWARTs web application currently supports the download of the full and partial results tables as a json or ascii file. An extra column is visible for option 3, indicating detectability of a possible kilonova associated with each galaxy. Additionally, the galaxies are plotted and a visibility plot is displayed for options 2 and 3. 

\section{Discussion}\label{Sec:5}

\subsection{User-testing and feedback}
 Additional functionalities that were requested from collaborators were implemented. For example, the option to download the galaxy list as an ascii file was added. The web application has been tested on Windows and Mac OS using leading browsers.

\subsection{Evaluation of execution time}
The mean execution time of the back-end galaxy retrieval algorithm was evaluated using cProfile for past GW events. Typical execution times are between 20--30 seconds, dependant upon the size (in MB) of the sky map FITS file, which in turn affects the time taken to download it. In addition to this execution time, 360 seconds is required to upload the new database and update the website on Heroku. 

\subsection{V0 algorithm}\label{section:evaluation}
V0 of the pre-processing algorithm was used throughout the first half of O3 (O3a).  This algorithm made use of \texttt{healpy} \citep{gorski2005healpix, Zonca2019} for sky map I/O and the \texttt{skimage} \citep{van2014scikit} \texttt{find\_contours} method was used to identify the contours enclosing 99\%, 90\%, and 50\% of the probability within the integrated probability map. Each contour was treated separately, by querying GLADE in Vizier using \texttt{Astropy} \citep{astropy:2018} with a circle enclosing each contour. The query was refined by only choosing galaxies within \texttt{DISTMEAN} $\pm$  \texttt{DISTSTD}. Galaxies were determined to be within the contour using \texttt{MOCPy}, a Python package for analysing Multi-Order Coverage (MOC) maps.  The coordinates, B magnitudes, and distances to these galaxies, as taken from the filtered GLADE V2 catalogue, were stored in an array to be used in subsequent probability calculations and the contour within which each galaxy lay was noted.

Throughout O3a, it became clear that some of the features of this algorithm could be optimised or improved. Occasionally some galaxies which were ranked highly on the 99\% list were not included in the 50\% or 90\% lists. This is due to high luminosity or distance probability measures, which therefore pushed galaxies outside the 50\% or 90\% regions up the 99\% list. For a fair comparison of galaxies, we choose to only evaluate the 99\% localisation region going forwards.

The range of galaxy distances considered is increased from \texttt{DISTMEAN} $\pm$ \texttt{DISTSTD} to \texttt{DISTMEAN} $\pm $ 5\,\texttt{DISTSTD} to ensure that galaxies which may be ranked highly due to B luminosity, but are outside of the 1$\sigma$ range, are included. This does not lengthen execution time and although it lengthens the galaxy lists, the majority of the extra galaxies included are ranked low due to their distances being significantly different from the mean distance in their pixels. 

The identification of each galaxy's contour is computationally expensive as each contour needs to be considered separately. The identification of contours was initially implemented for easy identification of the galaxies which lie in contours that are visible to the user. However, it is noted that the visibility option on the website can be used instead of this feature, therefore removing the identification of contour regions is sensible for the reduction of execution time. Additionally, the release of multi-order sky maps within gravitational wave alerts and the addition of a crossmatching function within the \texttt{ligo.skymap} package allow for quick crossmatching of a local galaxy catalogue with a sky map. We are implementing this function from now on to reduce execution time.

\subsection{Comparison to existing follow-up tools}
The NED Gravitational Wave Follow-up (GWF) Service\footnote[5]{The NASA/IPAC Extragalactic Database (NED) is operated by the Jet Propulsion Laboratory, California Institute of Technology, under contract with the National Aeronautics and Space Administration.} is a similar online tool which delivers a list of 2MASS galaxies within the 90\% localisation region of a GW event minutes after trigger. It ranks based on the 2MASS Redshift Survey (2MRS) Ks-band magnitude and does not make use of the information contained in the BAYESTAR/LALInference calculated probability and distance parameters. Throughout the start of O3, HOGWARTs galaxy lists have been broadly consistent with NED GWF galaxy lists. However, the NED GWF lists do not make a cut on distance, so provide all galaxies up to 200 Mpc. HOGWARTs lists also contain different candidate galaxies due to the use of a larger galaxy catalogue than NED. 

The Global Relay of Observatories Watching Transients Happen (GROWTH) Target of Opportunity (ToO) Marshal has been developed by the GROWTH collaboration to coordinate follow-up observations of multi-messenger transients \citep{kasliwal2019growth}. The Marshal responds to alerts by planning observations for a network of telescopes such as ZTF, Dark Energy Camera (DECam), and GROWTH-India and can send requests to robotic telescope queues. The tiling and scheduling features are beneficial for the network and for the wide field telescopes within that network, but the Marshal does not implement a galaxy identification or ranking feature. 

\subsection{HOGWARTs in O3}
HOGWARTs has responded to gravitational wave alerts in O3 and the website has been updated accordingly. In particular, HOGWARTs was used to conduct follow-up observations of the NS-BH merger S190814bv. 74 galaxies which were identified within the 99\% localisation region and ranked using the galaxy retrieval algorithm were followed up by the Telescopio Nazionale Galileo (TNG; \citet{TNG, TNG2}), the WHT \citep{WHT}, the Nordic Optical Telescope (NOT; \citet{NOT, NOT2}), the Gamma-Ray Burst Optical/Near-Infrared Detector (GROND; \citet{GROND}), and the Liverpool Telescope (LT; \citet{LT}). These follow-up observations took place within the ENGRAVE, GW@WHT, GROWTH, and GRAWITA collaborations. No candidate counterparts have yet been confirmed, however, the wide and deep coverage can place upper limits on electromagnetic counterparts from NS-BH mergers.

Each new NS-NS or NS-BH merger tests the galaxy-targeted strategy and the HOGWARTs system. It is therefore expected improvements will be made to the HOGWARTs system and perhaps additional functionality, such as scheduling and galaxy tiling tools, will be added to the current range of features available on the web application. However, it is currently considered that HOGWARTs is the first step in an individual's pipeline, allowing the individual to choose the scheduling that best suits their observatory. 

Throughout O3b it is expected that further improvements will be made to the system, including the possible transfer of the web application from Heroku to a private server to reduce the time it takes to update the website post-trigger. Further modifications to the distance limits are expected to consider the 3D credible region rather than the 2D credible region within distance limits. 

\section{Conclusions}
HOGWARTs is a web application for retrieving lists of candidate galaxies to observe in GW localisation regions. This publicly accessible tool contributes to the critical need for tools to assist astronomers in conducting GW follow-up.  This website is easily accessible worldwide and the code is open-source. Collaborations and single-users alike can make use of this tool to schedule their observations swiftly post-trigger and incorporate the website into their existing pipelines, tools, and processes. The back-end algorithm and website have been tested on past LVC sky maps and execution times have been analysed. With perhaps tens of NS-NS mergers expected during O3, there will be opportunities to test and improve this strategy and extend the functionality of the website over the coming months. 
\begin{acknowledgements}
LS acknowledges the Irish Research Council Postgraduate Scholarship No GOIPG/2017/1525. 

We are grateful to Alberto~Castro-Tirado, Morgan~Fraser, James~Gillanders, Kate~Maguire, Owen~McBrien,  Stephen~Smartt, and Brian~Van~Soelen for their useful feedback on the HOGWARTs web application.
We would like to thank the anonymous referee for their constructive comments that helped us improve the algorithm and content of the manuscript. 
\end{acknowledgements}

\bibliographystyle{aa}
\bibliography{mybibfile}

\end{document}